\documentclass{elsart}

\usepackage{amssymb}
\usepackage{graphicx}

\begin{document}

\begin{frontmatter}

\title{
On the Origin of Logarithmic-Normal Distributions: An analytical derivation, and its application to nucleation and growth processes
      }

\author{Ralf B.~Bergmann}
\address{Robert Bosch GmbH,
			Automotive Electronics, Quality Management Suppliers, Physical Analysis,
			P.O.Box 1342, 72703 Reutlingen, Germany}
\ead{Ralf.Bergmann@de.bosch.com, Phone: +49-7121-35-2998}
\author{Andreas Bill}
\address{California State University Long Beach,
			Dept. of Physics \& Astronomy,
			1250 Bellflower Blvd., Long Beach, CA 90840, USA}
\ead{abill@csulb.edu, Phone: +1-562-985-8616}

\begin{abstract}
The logarithmic-normal (lognormal) distribution is one of the most frequently observed distributions in nature and describes a large number of physical, biological and even sociological phenomena. The origin of this distribution is therefore of broad interest but a general derivation from basic principles is still lacking. Using random nucleation and growth to describe crystallization processes we derive the time development of grain size distributions. Our derivation provides, for the first time, an analytical expression of the size distribution in the form of a lognormal type distribution. We apply our results to the grain size distribution of solid phase crystallized Si-films.
 \end{abstract}

\begin{keyword}
A.1 Growth models, A.1 Nucleation, A.1 logarithmic normal distribution, A.1 random nucleation and growth
\PACS 81.10.Aj \sep 61.50.-f \sep 81.15.Aa
\end{keyword}

\end{frontmatter}

The logarithmic-normal (lognormal) size distribution is one of the most frequently encountered distributions in nature \cite{Limpert01,Aitchison69}. It has been shown early on that a large number of technical processes and even sociological phenomena such as income distributions \cite{Aitchison69} or the productivity of researchers \cite{Shockley57} follow a lognormal behavior. The distribution is also frequently observed as a result of various crystallization processes \cite{Bergmann97}; especially random nucleation and growth (RNG) processes usually result in a lognormal distribution of cluster or grain sizes \cite{Bergmann98}.
Lognormal size distributions in phase transformations have conventionally been attributed to coarsening \cite{Kurz80,Thompson88}. Our earlier work \cite{Bergmann98b}, however, pointed out that such distributions might occur due to time dependent kinetics of nucleation and growth, without the involvement of coarsening. Literature on nucleation describes the initial or early stages of nucleation \cite{Shi94,Kumomi99}  and there are a number of derivations of nucleation rates for these early stages, see eg. Ref.\cite{Schmelzer05}. A general description of the evolution of size distributions is e.g.~given in Ref.~\cite{Williams91} for aerosols and an overview on the development of size distributions in crystallization processes can be found in Ref.\cite{Kumomi02}. Furthermore, there are numerical approaches describing the size distributions e.g. during condensation of vapor \cite{Kozisek06}.\\
We provide for the first time an analytical derivation for the evolution of lognormal size distributions. Preconditions of our derivation are: i. The growth starts with the formation of  nuclei of a given critical size $g_c$, which is usually small as compared to the typical size $g$ of grains in the full-grown distribution. ii. Nuclei are formed on a random basis during simultaneous growth of other grains, which have already grown to a larger size g. iii. There is no coarsening (incorporation of smaller grains into larger ones.) These conditions apply to a large number of processes termed homogeneous {\it random nucleation and growth} (RNG). For the case of heterogeneous nucleation, the availability of nucleation sites has to be modified by a suitable extension of the present description. We first introduce a differential equation that describes the underlying physical concepts and present solutions for physically relevant cases. We then apply our results to the grain size distribution of solid phase crystallized Si-films.

We start by establishing the differential equation containing the essential ingredients of a RNG-process.
The formation of nuclei requires a critical size $g_c$. Therefore, we describe the contribution of nucleation to the grain-size distribution $N$ at time $t$ by
\begin{equation} 
\label{Eq1}
\frac{\partial N(g,t)}{\partial t} = I(t)\, \delta\left(g - g_c\right),
\end{equation}
where $g$ is the diameter of grains, $I(t)$ is the time dependent nucleation rate and $\delta$ is the Dirac delta-function. Once nuclei are formed, the change of grain-size distribution $N$ in time is due to the growth of nuclei and is obtained from the continuity equation
\begin{equation} 
\label{Eq2}
\frac{\partial N(g,t)}{\partial t} = - \mathbf{\nabla}\cdot \mathbf{j}
\end{equation}
with the particle current density $\mathbf{j} = N(g,t)\, \mathbf{v}(\mathbf{g},t)$. We now assume that the growth rate  $\mathbf{v}(\mathbf{g},t)$ is isotropic and independent of $g$. Using polar or spherical coordinates depending on whether the crystallization of a thin film or a bulk material is considered, all angle dependent terms vanish and we obtain
\begin{equation} 
\label{Eq3}
\mathbf{\nabla}\cdot \mathbf{j}
=
\frac{v(t)}{g^{d-1}} \frac{\partial \left(g^{d-1}N(g,t)\right)}{\partial g}.
\end{equation}
This expression holds for dimensions $d = 1, 2$ and 3.
\footnote{Note, that in contrast to previous work \cite{Bergmann98b}, the contribution from the divergence is written in terms of the diameter $g$ of the grain rather than its radius $r$. Consequently, the growth-rate $v(t)$ is twice the growth rate in Ref.~\cite{Bergmann98b}.} Introducing Eq.~(\ref{Eq3}) into the continuity Eq.~(\ref{Eq2}) and combining Eq.~(\ref{Eq1}) with Eq.(\ref{Eq2}), one obtains the nucleation and growth equation in $d$ dimensions
\begin{equation} 
\label{Eq4}
\frac{\partial N(g,t)}{\partial t}
=
I(t) \delta\left(g - g_c\right) - \frac{v(t)}{g^{d-1}} \frac{\partial \left(g^{d-1}N(g,t)\right)}{\partial g}.
\end{equation}
Since the increase in diameter $g$ is determined by the time dependent growth rate $v(t)$, the grain size for $t\to\infty$ may simply be derived from $dg(t) = v(t) dt$ and results in 
\begin{equation} 
\label{Eq5}
g_{inf} = g_c + \int_0^\infty v(t) dt.
\end{equation}
We define the quantity $\tilde{N}(g,t) = g^{d-1}N(g,t)$ in order to transform Eq.~(\ref{Eq4}) into the somewhat simpler shape
\begin{equation} 
\label{Eq6}
\frac{\partial \tilde{N}(g,t)}{\partial t} + v(t) \frac{\partial\tilde{N}(g,t)}{\partial g}
=
g^{d-1}\,I(t)\, \delta\left(g - g_c\right).
\end{equation}
In order to solve this partial differential equation, we use the well-known Laplace transform and obtain an ordinary differential equation that can be solved analytically. Applying the inverse Laplace transform to the solution leads to the final result. The calculations involved in this procedure are beyond the scope of this contribution and will be described separately \cite{Bill08}.\\
In order to obtain specific solutions of Eq.~(\ref{Eq4}) one has to make a reasonable choice for the time dependence of nucleation and growth rates $I(t)$ and $v(t)$. As RNG proceeds, nucleation is decreasing due to the decreasing fraction of available material, while the average growth rate decreases due to impingement of neighbouring grains. The time dependent fraction of material $X(t)$ crystallized during RNG-processes is frequently described by the Avrami-Mehl-Johnson (AMJ) \cite{Schmelzer05} expression
\begin{equation} 
\label{Eq8}
X_d(t) =
\left\{1- \exp\left[-\left(\frac{t-t_0}{t_c}\right)^{d+1}\right]\right\} \Theta\left(\frac{t-t_0}{t_c}\right)
\end{equation}
with the critical crystallization time $t_c$, the incubation time $t_0$ preceding crystallization, and the dimensionality of the growth process $d = 2$ for film growth and $d = 3$ for bulk growth. $\Theta(t)$ is the Heaviside function. For an experimental setup to detect the crystallization kinetics of thin films, see e.g.~Ref.~\cite{Bergmann96}. The fraction of material not crystallized at time $t$ and thus still available for further nucleation and growth is consequently given by
\begin{equation} 
\label{Eq9}
Y_d(t) = 1- X_d(t) =
\exp\left\{-\left(\frac{t-t_0}{t_c}\right)^{d+1}\right\} \Theta\left(\frac{t-t_0}{t_c}\right)
\end{equation}
In the following we will assume $t_0 = 0$, as the present paper does not address incubation effects.
Following the above discussion, crystallization occurs through nucleation and growth processes obeying different time dependencies. Accordingly, we introduce two critical time constants $t_{cI}$ for nucleation and $t_{cv}$ for growth. We also consider different, though similar, functional forms for the nucleation rate
\begin{equation} 
\label{Eq10}
I_n(t)  = I_0 Y_n^I(t) = I_0 \exp\left[- \left( \frac{t}{t_{cI}} \right)^{n+1} \right]
\end{equation}
and the growth rate
\begin{equation} 
\label{Eq11}
v_m(t)  = v_0 Y_m^v(t) = v_0 \exp\left[- \left( \frac{t}{t_{cv}} \right)^{m+1} \right],
\end{equation}
and assume $m \neq n$ in order to account for the different processes at work.
We have found the complete analytical solution of Eq.~(\ref{Eq4}) with the above rates. Combinations of $n = 1, 2, 3$ and $m = 1, 2$ leads to a variety of functional forms that will be described elsewhere \cite{Bill08}. Assuming that nucleation is limited by the available material for crystallization, we identify $n$ with the dimension of the growth process, and therefore use $n = d$. The choice of $m$ is, on the other hand, less obvious. Choosing $m=0$ results $i)$ in a lognormal type distribution, and $ii)$ in a more accurate description of the experimental results discussed below. The correspondence of our theoretical result with experimental data of a specific RNG process may therefore allow to give an independent determination of the appropriate value of $m$, and thereby of the time dependence of the growth rate $v(t)$ in the model.  In general, $v_m(t)$ will depend on the details of the process considered. A description of such details (e.g.~the effect of hard or soft impingement of grains) is, however, beyond the scope of the present work.\\
Solving Eq.~(\ref{Eq4}) with the rates given in Eqs.~(\ref{Eq10},\ref{Eq11}) we obtain for the case $m = 0$
\begin{eqnarray} 
\label{Eq12}
N(g,t) &=& 
\left(\frac{I_0}{v_0} \right) \left(\frac{g_c}{g}\right)^{d-1} \frac{1}{\alpha}
\exp\left[ (-1)^n\left( \frac{t_{cv}}{t_{cI}} \ln \alpha \right)^{n+1} \right]\nonumber \\
&&\hspace*{3cm}\times
\left[ \Theta\left(\frac{g-g_c}{t_{cv}v_0}\right) - \Theta\left(\frac{g-g_{max}(t)}{t_{cv}v_0}\right) \right],
\end{eqnarray}
with
\begin{equation} 
\label{Eq13}
\alpha = \frac{g-g_c}{t_{cv}v_0} + e^{-t/t_{cv}}.
\end{equation}
and the maximal grain size given by Eq.~(\ref{Eq5})
\begin{equation} 
\label{Eq14}
g_{max}(t) = g_c + t_{cv} v_0 \left(1- e^{-t/t_{cv}}\right)
\end{equation}
Eq.~(\ref{Eq12}) is a remarkable result in that it displays a mathematical structure close to the lognormal distribution, a fact not previously recognized in the literature. This is seen more clearly when considering the final distribution obtained once the nucleation and growth processes have been completed ($t\to\infty$). Indeed, we then obtain from Eq.~(\ref{Eq12})
\begin{eqnarray} 
\label{Eq15}
N(g) &=&
\left(\frac{g_c}{g}\right)^{d-1} \frac{I_0t_{cv}}{g-g_c}
\exp\left[(-1)^d\left(\frac{t_{cv}}{t_{cI}}\ln\left(\frac{g-g_c}{t_{cv}v_0}\right)\right)^{d+1}\right]\nonumber\\
&&\hspace*{3cm}\times
\left[ \Theta\left(\frac{g-g_c}{t_{cv}v_0}\right) - \Theta\left(\frac{g-g_{inf}}{t_{cv}v_0}\right) \right],
\end{eqnarray}
with
\begin{equation} 
\label{Eq16}
g_{inf} = \lim_{t\to\infty} g_{max}(t) = g_c + t_{cv}v_0.
\end{equation}
For $g \gg g_c$ we can further simplify the result to
\begin{equation} 
\label{Eq17}
N_{>}(g) =
\left(\frac{I_0}{v_0}\right)\left(\frac{g_{inf}}{g_c}\right) \left(\frac{g_c}{g}\right)^{d}
\exp\left[ (-1)^d\left(\frac{\ln\left(\frac{g}{g_{inf}}\right)}{\frac{t_{cI}}{t_{cv}}}\right)^{d+1} \right]
\times
\Theta\left(\frac{g_{inf}-g}{t_{cv}v_0}\right),
\end{equation}
and
\begin{equation} 
\label{Eq18}
g_{inf} = v_0 t_{cv}.
\end{equation}
A comparison of the above result with the lognormal distribution
\begin{equation} 
\label{Eq19}
\Lambda(g) = \frac{1}{\sqrt{2\pi} \sigma g} \exp\left[ - \frac{1}{2} \left( \frac{\ln\left(g/M\right)}{\sigma} \right)^2 \right]
\end{equation}
with the median $M$ and the width $\sigma$ shows that we obtain a size distribution that differs from the lognormal distribution in three respects: a) there is an additional dimension-dependent prefactor $1/g^{d-1}$, b) the exponent of the lognormal distribution also depends on the dimensionality of the growth process and c) the distribution is multiplied by a cut-off function at the maximal grain size ($g_{inf}$ at $t\to\infty$). Note that Eq.~(\ref{Eq17}) contains the ratio $(g_c/g)^d$ which is directly proportional to the area or volume of grains. Due to the differences between expressions (\ref{Eq17}) and (\ref{Eq19}), we cannot directly identify the parameter $M$ with $g_{inf}$  or $t_{cI}/t_{cv}$ with $\sigma$.\\
Although differential equations similar to the one introduced here have been discussed in the literature in various fields (see, e.g.~Ref.~\cite{Williams91}, \cite{Kumomi02} and references therein), our solution departs from those found elsewhere in several respects. One of them is the presence of the cutoffs at low and high grain sizes in Eqs.~(\ref{Eq12}),(\ref{Eq15}), (\ref{Eq17}), which directly result from our derivation and the proper treatment of the Dirac distribution. Their appearance correctly reproduces the physics of the RNG process. We also note that in the case $n = 1$, $m = 0$ the particular form of the rates $I_n(t)$ and $v_m(t)$ exactly leads to the lognormal distribution generally introduced empirically to fit data. To the best of our knowledge this is the first analytical derivation of a lognormal type distribution reported so far.

We now apply our result to experimental data obtained from the solid phase crystallization of amorphous Si as an example for an RNG-process. The experimentally determined size distribution is taken from Ref.~\cite{Bergmann98}. The corresponding parameter $v_0 = 1.72 \pm 0.34$ $\mu$m/h is taken from Ref.~\cite{Bergmann98b}, which also states the crystallization parameters $t_0 =1.5 \pm 0.5$h and $t_c = 4 \pm 0.5$h, necessary for Eq.~(\ref{Eq9}), using $d = 2$ for amorphous Si films solid phase crystallized at 600$^\circ$C. The critical grain size $g_c$ is of the order of a few nanometers. Since a reliable experimental value of $I_0$ is not available but $I_0$ and $g_c$ only appear as a product in the size distribution, we treat the constant, grain size-independent prefactor of Eq.~(\ref{Eq17}) as a fit parameter. Note that the use of experimental values of $I_0$ would have to take into account a factor related to the transformation to polar or spherical coordinates when introduced in the solutions described above \cite{Bill08}.
\unitlength1cm
\begin{figure}
\begin{center}
\includegraphics*[scale=0.9]{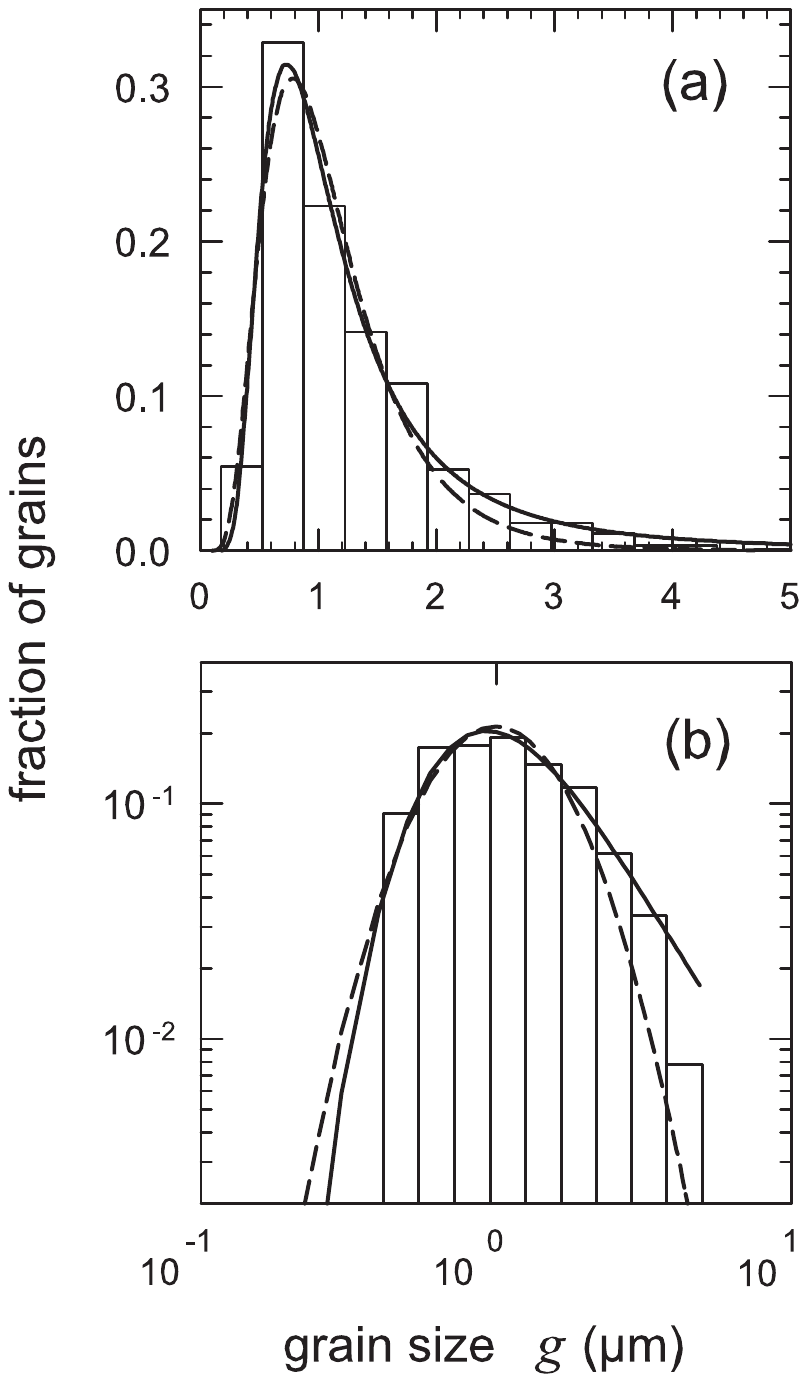}
\end{center}
\vspace*{-10cm}
\caption{Grain size distribution (histogram) of a Si-film with a thickness of 1 $\mu$m solid-phase crystallized at 600¡C for 15h. Fit to Eq.~(\ref{Eq17}) (solid line) and to the lognormal distribution given in Eq.~(\ref{Eq19}) (dashed line) for comparison. a) Distribution represented on a linear size scale, b) Distribution represented on a double logarithmic scale.}
\label{Fig1}
\end{figure}

Figure 1 compares the grain size distribution of a fully crystallized Si film to our calculations above. Fig. 1a shows a histogram of the distribution of an ensemble of about 900 grains. Fig. 1b depicts the same set of data, however the histogram is created using the distribution of $\log_{10} g$ of the individual grains. Every bar of the original distribution of Fig. 1a represents a number of $N(g)dg$ grains in the size interval $[g,g+dg]$. A transformation of a distribution function $N(g)$ to a log-distribution scale is written $N(g)dg = g\, N(\ln g)d\ln g$. This procedure enables a better visualization of the distribution of small and less frequently occurring grains and is applied to the size distribution functions derived above \footnote{The same procedure has been used for Fig.~1 of Ref.~\cite{Bergmann98b}. Note that this transformation changes the shape of the distribution: while the measured grain-size distribution with a crystallized fraction of 2.8\% is indeed non-monotonic, the calculated distribution is not. Although both distributions appear non-monotonic on the logarithmic size scale, the calculated distribution only becomes non-monotonic once the crystallized fraction exceeds about 16\%. Nevertheless, the validity of the concept presented in Ref.~\cite{Bergmann98b} is now supported by the generic derivation given here.}

Fitting the lognormal distribution of Eq.~(\ref{Eq19}) (dashed line) to the experimentally observed distribution gives $M = 1 \pm 0.04 \mu$m, $\sigma = 0.49 \pm 0.03$. For a fit of the experimental data to Eq.~(\ref{Eq17}) (solid line), we use $d = 3$ and $g_{inf} = 4.9 \mu$m, the largest grain size measured in this distribution. The best fit results in $t_{cI} / t_{cv}= 1.74 \pm 0.02$. Using, as stated above, a growth rate of $v_0 = 1.72 \pm 0.34 \mu$m/h one obtains the following reasonable values $t_{cv}= 2.9 \pm 0.6$h and $t_{cI} = 5 \pm 1$h. It should be noted that $t_{cv}$ and $v_0$ can be varied within reasonable limits as long as the ratio $t_{cI}/t_{cv}$ and $g_{inf}$ remain constant. A fit using $d = 2$ was also attempted, but the result is of inferior quality. We attribute this finding to the fact that a large fraction of grains have a diameter smaller than the film thickness of $1 \mu$m and growth is therefore mainly three-dimensional. Parameters obtained from other crystallized samples lead to comparable results. A comparison to further experimental results including partially crystallized samples will be presented elsewhere \cite{Bill08}. We stress the fact that, contrary to the fit using the lognormal distribution Eq.~(\ref{Eq19}) (dashed line on Fig.1), our fit (solid line) is based on a physical model and theoretical derivation of Eq.~(\ref{Eq17}) applied to RNG processes.

In conclusion, we present a generic differential equation for the time-dependant grain size distribution resulting from random nucleation and growth (RNG) processes. We specify the nucleation and growth rates according to the well established Avrami-Mehl-Johnson-Equation, and for the first time achieve an analytical derivation of logarithmic normal-type distributions. We therefore propose that the frequently observed lognormal size distribution resulting from RNG originates from the dynamics of the process as described here. Our comparatively simple approach is well supported by the grain size distribution found in fully crystallized Si-films. Our model may therefore significantly benefit the understanding and tailoring of technically relevant nucleation and growth processes.

\newpage
\ack{A.B.~thanks the SCAC at CSU Long Beach and the Research Corporation for support}

{\bf Note added in proof}\\[.3cm]
Recently, we became aware of an early, rarely cited paper of Kolmogorov in Dokl.Akad.Nauk. SSR {\bf 31}, 99 (1941). Using a statistical approach, the author derived the logarithmic normal distribution for particle sizes resulting from grinding. Our forthcoming paper \cite{Bill08} will discuss the relevance of this approach and the resulting distribution of grain sizes obtained.


\end{document}